\documentclass[12pt]{article}
\usepackage{epsfig}
\newlength{\defbaselineskip}
\newcommand{\setlinespacing}[1]%
           {\setlength{\baselineskip}{#1 \defbaselineskip}}

\begin{document}

\title{The study of the nucleus-nucleus interaction potential for $^{16}$O+$^{27}$Al and $^{16}$O+$^{28}$Si fusion reactions}

\author{O. N. Ghodsi, R. Gharaei\thanks{Email: r.gharaei@stu.umz.ac.ir} \\
\\
{\small {\em  sciences Faculty, Department of Physics, University of Mazandaran}}\\
{\small {\em P. O. Box 47415-416, Babolsar, Iran}}\\
}
\date{}
\maketitle

\begin{abstract}
\noindent Using the Monte Carlo simulation method accompanied by the
modifying effects of the density distributions overlapping, we have
examined the nuclear matter incompressibility effects for asymmetric
systems with light nuclei, namely $^{16}$O+$^{27}$Al and
$^{16}$O+$^{28}$Si fusion reactions. The obtained results show that
the nuclear equation of state has considerable influence on the
calculation of fusion probabilities for these asymmetric systems.

\end{abstract}

\newpage
\setlinespacing{1}
{\noindent \bf{1. INTRODUCTION}}\\

\noindent The recent fusion experiments using Medium-Heavy nuclei
show remarkable changes in the fusion excitation functions (FeF) at
deep sub-barrier energies [1, 2]. Further research indicated that
these changes can be attributed to the hindrance effects in fusion
reactions [3]. In theoretical models which are based on the sudden
approximation to calculate the nuclear potential, one can explain
the observed changes in slope of the FeF by using the simulation of
repulsive core effects. In fact these modifications take into
account the effects of Pauli Exclusion Principle in calculation of
nuclear potential and correct it in the inner regions [4, 5]. In
general, the shallow potential resulting from the modified M3Y
interaction with repulsive core can explain fusion cross sections at
deep sub-barrier energies for symmetric (asymmetric) systems with
positive Q-value [4]. This research has been done on symmetric
Medium-Medium nuclei fusion reactions such as $^{58}$Ni+$^{58}$Ni
and $^{64}$Ni+$^{64}$Ni [4, 6], Light-Heavy nuclei such as
$^{16}$O+$^{208}$Pb [5] and symmetric Light-Light nuclei such as
$^{16}$O+$^{16}$O [7] successfully. So we have motivated to
investigate the hindrance effects on asymmetric Light-Light
reactions at energies above the fusion barrier. For this purpose we
have selected $^{16}$O+$^{27}$Al and $^{16}$O+$^{28}$Si reactions
which their Q-values for producing the compound nucleus are
positive.

In Sec. 2 we shall briefly discuss about the employed model in the
calculation of total interaction potential and correction of this
model by using an additional repulsive force in the form of a
repulsive core for modeling the hindrance effects. Calculation
process of potential and fusion cross section for selected reactions
is given in Sec. 3. Section 4 is devoted to some concluding remarks.
\\
\\
\\

\noindent{\bf {2. THEORY}}\\
\\
\noindent{\bf {A. Internuclear Potential}}\\

The interaction potential between two nuclei is given by,

\begin{equation} \label{1}
V(r)=V_{C}(r) + V_{N}(r),
\end{equation}
where $V_{C}(r)$ and $V_{N}(r)$ are resulting from the large-range
Coulomb repulsion and short-range nuclear attraction in
nucleon-nucleon (NN) interactions, respectively. The nuclear
potential between target and projectile nuclei plays a crucial role
in the study of fusion cross section.  In order to calculate the
nuclear part of total potential we have employed a simulation method
that it has reported in our previous work [8]. In this simulation
technique, we have used the neutrons and protons densities for
calculation of nuclear potential. This capability allows us to
consider the effects of the surface nucleons in the calculation of
interaction potential [9].
\\
\\
\\
\noindent{\bf {B. Modification of Internuclear Potential}}\\

Double folding (DF) model, because of using sudden approximation,
predicts negative values for inner regions of the total potential
(see Fig. 1). When two interacting nuclei begin to overlap and
nuclear matter density roughly becomes twice the saturation density
($\rho\approx2\rho_{0}$), it could be predicted that an extra
repulsive force prevents the compressing of nucleus. In order to
consider the modifications of the incompressibility effect on the
calculation of the nuclear potential, it is suggested that this can
be simulated by adding a zero-range interaction as the following
form

\begin{equation} \label{2}
\upsilon_{rep}(\mathbf{s})=V_{rep}\delta(\mathbf{s}),
\end{equation}
to the nucleon-nucleon interaction \cite{6}, where $V_{rep}$ is a
constant parameter. In fact, this interaction simulates the Pauli
Exclusion Principle effect on the calculation of nuclear potential.
The constants of this repulsive force can be calculated by using the
nuclear part of the total potential when two interacting nuclei
overlap completely. It has been shown that the value of $V_{N}$ at
$r=0$ can be estimated using the following relation [6],

\begin{equation} \label{3}
V_{N}(r=0)=\Delta{V}\approx\frac{A_{P}}{9}K,
\end{equation}
where $A_{P}$ is the mass number of smallest nucleus in the case of
an asymmetric system and $K$ is the incompressibility constant and
is given by the following relation,

\begin{equation} \label{4}
K=9\bigg(\rho^{2}\frac{\partial^{2}\varepsilon}{\partial\rho^{2}}\bigg)_{\rho=\rho_{0}}.
\end{equation}
where $\varepsilon(\rho)$ is the binding energy per particle of
nuclear matter. We have used Thomas-Fermi model to calculate
$\varepsilon(\rho)$ [10]. The nuclear part of total potential can
also be calculated using the following relation,

\begin{equation} \label{5}
V_{N}(r=0)=V_{dir}(r=0)+V_{exc}(r=0)+\int d\mathbf{r}_1\int
d\mathbf{r}_2\rho_1(\mathbf{r}_1)V_{rep}\delta(\mathbf{r}_{12})\rho_2(\mathbf{r}_2),
\end{equation}
where $V_{exc}$ and $V_{dir}$ are direct and exchange parts of
nuclear potential due to the M3Y force. In computation of the
integral part of Eq. (5), the parameter $V_{rep}$ and the
diffuseness constant of density distributions of target and
projectile nuclei, assuming $a_{T}=a_{P}=a_{rep}$, adjusted such
that the calculated value of nuclear potential at $r=0$ and the
height of the fusion barrier are in the agreement with the predicted
values from the Eq. (3) and the corresponding experimental data for
each reaction, respectively.
\\
\\
\\
\noindent{\bf {C. One-Dimensional Barrier Penetration Model (ODBPM) }}\\

To study of fusion cross section, we have used the ODBPM [11, 12].
In This formalism, the cross section for complete fusion is given by

\begin{equation} \label{6}
\sigma_{fus}(E)=\sum^{\infty}_{\ell=0}\sigma_{\ell}(E),
\end{equation}
where the partial-wave cross sections can be calculated using

\begin{equation} \label{7}
\sigma_{\ell}(E)=\frac{\pi\hbar^{2}}{2\mu E}(2\ell+1)T_{\ell}(E).
\end{equation}

In this relation, $\mu$ is the reduced mass of the projectile and
target systems and $T_{\ell}(E)$ is the transmission coefficient for
angular momentum $\ell$ through potential barrier at center-of-mass
energy $E$. This latter coefficient can be computed using the WKB
[13] approximation for penetration through the barrier,

\begin{equation} \label{8}
T_{\ell}(E)=\bigg[1+\textmd{exp}\bigg(2\sqrt{\frac{2\mu}{\hbar^2}}\int_{r_{1\ell}}^{r_{2\ell}}dr[V_{0}(r)+\frac{\hbar^2\ell(\ell+1)}{2\mu
r^2}-E]^{1/2}\bigg)\bigg]^{-1},
\end{equation}
where $r_{1\ell}$ and $r_{2\ell}$ are classical turning points for
angular momentum $\ell$ and $V_{0}(r)$ is total potential for
$\ell=0$. We refer the reader to Ref. [11-13], where this formalism
is completely explained.
\\
\\
\\
\noindent{\bf {3. CALCULATIONS}}\\

For fusion reactions with light interaction nuclei, because the
radius of the target and projectile nuclei are small, one expects
that the nuclear matter incompressibility effects have less
importance in the calculation of total potential, particularly in
place of forming Coulomb barrier. In order to investigate the effect
of nuclear matter incompressibility on light fusion reactions, we
have selected two systems $^{16}$O + $^{27}$Al and $^{16}$O +
$^{28}$Si. Because mass numbers of target and projectile nuclei in
each of these reactions are smaller than or equal to 28 and also
their Q-values are positive (see Table 2). In the beginning, we have
calculated the total potential for the above reactions using the
Monte Carlo simulation method and NN interaction of M3Y-Paris type
\cite{8}. The density distributions of target and projectile nuclei
are parameterized by using the two-parameter Fermi-Dirac
distribution functions such that,

\begin{equation} \label{9}
\rho_i(r) = \frac{\rho_0}{1+\exp{}[(r-R_{0i})/a]}.
\end{equation}
The values of diffuseness $a_{o(n,p)}$ and radius $R_{o(n,p)}$
parameters for proton and neutron densities, that are obtained by
Hartree-Fock-Bogoliubov (HFB) calculations [14], have been listed in
Table 1. The calculated total potentials for $^{16}$O + $^{27}$Al
and $^{16}$O + $^{28}$Si reactions have been shown in Fig. 1. As a
result, the nuclear potential using the M3Y nucleon-nucleon
interaction results the negative values for the inner part of the
total potential. With adding the repulsive potential to the NN
interaction one can take into account the effects of the nuclear
matter incompressibility on the nuclear potential calculation. This
correction is performed using the method introduced in the previous
section and its results are shown in Fig. 1 by M3Y+repulsion
potential. Our manner for calculation of parameters used in
repulsive core modeling is the fusion cross sections resulting from
the M3Y+rep potential have a good agreement with their corresponding
experimental data [15, 16]. The obtained values for these
coefficients and incompressibility constant $K$ as well as the
values of shallow pocket, $V_{pocket}$, for each reaction have been
listed in Table 2. As it is shown in Fig. 1, modifying of M3Y
potential leads to appearance of shallow pocket in the inner regions
of the total potential and consequently improves the fusion cross
section agreement with experimental data (see Fig. 2). Since our aim
in this paper is investigation of the compressibility effects on the
calculation of fusion cross sections for Light-Light nuclei at
energies above the fusion barrier, we have ignored the
coupled-channel effects in the calculation of cross sections.
\\
\\
\\
\noindent{\bf {4. RESULT AND DISCUSSION}}\\

In this paper we have investigated the dependence of the total
interaction potential on the nuclear equation of state for the
fusion reactions with asymmetric Light-Light nuclei. For this
purpose, we have calculated the nuclear potential for $^{16}$O +
$^{27}$Al and $^{16}$O + $^{28}$Si systems using the Monte Carlo
method and M3Y nucleon-nucleon force and modified M3Y+DF potential
(M3Y+repulsion). The obtained results are shown in Fig. 1, where the
vertical lines show the distance that total overlapping density
increases roughly that of normal matter, i.e., $\rho_{0}= 0.161$
fm$^{-3}$. As one can see, considering the modifying effects of
nuclear matter incompressibility on the calculation of total
potential causes increase of the height and thickness of Coulomb
barrier as well as appearing shallow pocket in the inner regions of
interaction potential. In present work, we have used the ODBPM for
computation of the fusion cross section. As a result, the increase
of thickness of fusion barrier causes decreasing of penetration
coefficient and consequently fusion cross sections. It is evident
from Fig. 2 that the obtained results for fusion cross sections
resulting from the M3Y+repulsion potential, in comparison with the
M3Y potential, are in the better agreement with the corresponding
experimental data in above energies. This subject shows that in
fusion reactions with light target and projectile nuclei, in spite
of the littleness of interaction nuclei radius in comparison with
heavy-nuclei, nuclear equation of state can be important in study of
light fusion reactions.
\\
\\

\newpage
Table 1. The values of the diffuseness $a_{o(p,n)}$ and radius
$R_{o(p,n)}$ parameters for proton and neutron density distributions
of $^{16}$O, $^{27}$Al and $^{28}$Si nuclei which determined by HFB
calculation \cite{14}.

\begin{center}
\begin{tabular}{c c c c c}
  \hline
  \hline
  Nucleus   &     $R_{0p}$ (fm)   &    $a_{0p}$ (fm)   &    $R_{0n}$ (fm)   &    $a_{0n}$ (fm) \\
  \hline
  $^{16}$O      &   2.6986   &   0.4469     &  2.6519        &   0.4602      \\
  $^{27}$Al     &   3.1595   &   0.4646     &  3.1361        &   0.4782      \\
  $^{28}$Si     &   3.1984   &   0.4750     &  3.1671        &   0.4726      \\
  \hline
\end{tabular}
\end{center}

Table 2. The obtained results for $V_{rep}$, $a_{rep}$, $K$ and
$V_{pocket}$ for $^{16}$O + $^{27}$Al and $^{16}$O + $^{28}$Si reactions.
The Q-values of reactions for the production of the compound nucleus
are listed in the last column.

\begin{center}
\begin{tabular}{c c c c c c}
  \hline
  \hline
  Reaction   &     $a_{rep}$ (fm)   &    $V_{rep}$ (MeV.fm$^{3}$)   &    $K$ (MeV)   &    $V_{pocket}$ (MeV)   &   Q-value (MeV) \\
  \hline
  $^{16}$O+$^{27}$Al      &   0.325   &   445.6     &  234.22     &   0.48   &  +14.254      \\
  $^{16}$O+$^{28}$Si      &   0.340   &   454.0     &  234.44     &   3.94   &  +11.318      \\
  \hline
\end{tabular}
\end{center}


\newpage
\begin{figure}
\begin{center}
\includegraphics{Fig.}
\end{center}
\vspace{15cm} \caption{} Fig. 1 The calculated total potentials for
(a) $^{16}$O+$^{27}$Al and (b) $^{16}$O+$^{28}$Si fusion
reactions. The dashed and dotted curves are based on the M3Y and
M3Y+repulsion potential, respectively. Vertical line shows the place
where the nuclear matter density due to the overlapping of
interacting nuclei increases to values about twice the nucleon
matter saturation density.
\end{figure}


\newpage
\begin{figure}
\begin{center}
\includegraphics{Fig.}
\end{center}
\vspace{15cm} \caption{} Fig. 2 The calculated complete fusion cross
sections for (a) $^{16}$O+$^{27}$Al and (b) $^{16}$O+$^{28}$Si
fusion reactions. The experimental data have been extracted from
Refs. [15] and [16] for $^{16}$O+$^{27}$Al and $^{16}$O+$^{28}$Si
systems, respectively .
\end{figure}


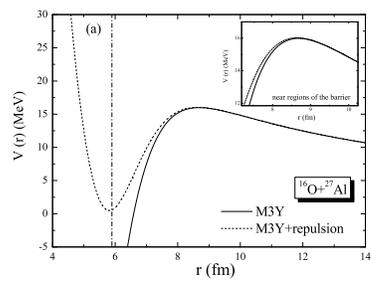
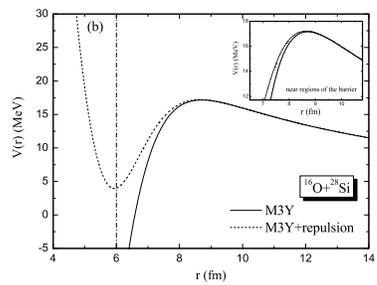
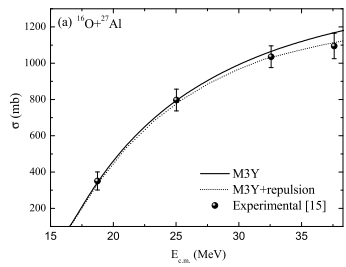
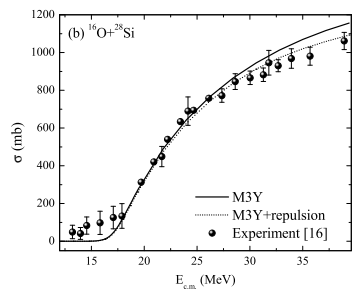
\end{document}